\documentclass[onecolumn]{svjour3}

\newcommand\T{\rule{0pt}{3.6ex}}

\usepackage{amsmath}
\usepackage{amssymb}
\usepackage{amscd}		

\usepackage{graphicx}

\usepackage{mathptmx}

\newcommand{\beqa}{\begin{eqnarray}}
\newcommand{\eeqa}{\end{eqnarray}}

\def\jpb#1{{ J.\ Phys.\ B} {\bf#1}}
\def\jpa#1{{ J.\ Phys.\ A} {\bf#1}}

\def\pra#1{{ Phys.\ Rev. A\/} {\bf#1}}
\def\prb#1{{ Phys.\ Rev. B\/} {\bf#1}}
\def\pre#1{{ Phys.\ Rev. E\/} {\bf#1}}
\def\prl#1{{ Phys.\ Rev.\ Lett.} {\bf#1}}
\def\sci#1{{ Science} {\bf#1}}

\newcommand{\beq}{\begin{equation}}
\newcommand{\eeq}{\end{equation}}

\journalname{Quantum Information Processing}

\begin{document}

\title{Modulated Entanglement Evolution Via Correlated Noises}

\author{Brittany Corn$^\dag$ \and Ting Yu$^*$}

\institute{Department of Physics and Engineering Physics, Stevens Institute of Technology, 
Hoboken, NJ 07030, USA\\
Emails: $^\dag$bcorn@stevens.edu,  $^*$ting.yu@stevens.edu }

\date{Received:  July 28,  2009}

\maketitle

\begin{abstract} 
We study entanglement dynamics in the presence of correlated environmental 
noises. Specifically, we investigate the quantum entanglement dynamics of 
two spins in the presence of correlated classical white noises, deriving Markov master
equation and obtaining explicit solutions for several interesting classes of initial states including
Bell states and X form density matrices.  We show how entanglement can be enhanced or 
reduced by the correlation between the two participating noises.

\end{abstract}

\section{Introduction}
Many proposed applications in quantum computing \cite{Nielson}, quantum communication \cite{Bennett}, and quantum 
cryptography \cite{BennettBrassard} revolve 
around harnessing the inherent correlation between quantum particles, called entanglement \cite{Polishfamily}. Although
 quantum mechanics dictates that these 
coherence effects are intrinsic in certain systems, even when the atoms or particles are non-local, there is an overall 
weakening due to coupling 
to noisy environments \cite{Zurek} that eventually leads to the fast decay of entanglement \cite{Hfamily,Simon-Kempe02,Yu-Eberly02,Dur-Briegel04,Buchleitner}, as in the case of amplitude or phase noise, or 
even the sudden death of entanglement (ESD) in the worst scenarios \cite{Yu-EberlySci09,Davidovich-etal07,Eberly-YuSci07,Yu-EberlyPRL04,Yu-EberlyPRL06}. 
As such, the study of the controlled entanglement dynamics is of much importance to the prospects of maintaining quantum information \cite{control20091,FicTan06,ASH,zeno,control2009,control20092,control1,control2}.
Moreover, model systems that theoretically exhibit  the rebirth of entanglement have been proposed and discussed in several cases \cite{Yonac2007,birth0,birth1,birth2}. 
\begin{figure}[b!]
\begin{center}
\includegraphics[width=5.0cm]{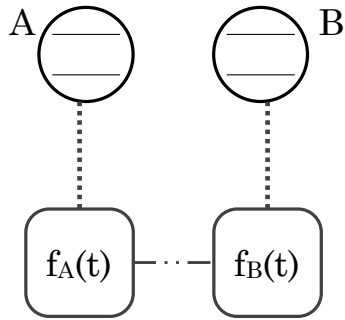}
\caption{The model is composed of two uncoupled qubits A and B individually coupled to respective  stochastic fields, $f_A(t)$ and $f_B(t)$, which are then correlated. }
\end{center}
\end{figure}
	
In this paper, we consider a model system consisting of two uncoupled qubits A and B  interacting with stochastic fields, $f_A(t)$ and $f_B(t)$, respectively, shown in Fig.1. 
The setup of the system is quite similar to the two-qubit local dephasing channel of \cite{Yu-Eberly02} in which the qubits were found to disentangle in a shorter time 
than their individual atom-field local dephasing times. On the other hand, if the two qubits are coupled to a common environment, then entanglement was shown to
be preserved for a class of initial states living in a subspace called decoherence-free subspace (DFS) \cite{dfs,dfs1}.  By introducing a correlation between the two noisy sources, we 
can show that the entanglement between the two qubits can be enhanced or reduced by properly choosing the correlation of the two participating external
 noises.  
 
The paper is organized as follows:  We present the specifics of the model system in Section II, which leads to deriving the master equation in the Markov regimes.
The entanglement evolution under the correlated noisy sources are studied in Section III. We have shown that dependent on initial states the correlation between
 two external noises can either enhance or reduce the existing entanglement of the qubit systems. We conclude in Section IV.

\section{Qubit Model and the Master Equation}

Our model consists of two identical, separated qubits (two-level atoms, spins, excitons, etc.) each having transition frequency $\omega$ and each coupled to 
a separate classical random field, $f_A(t)$ and $f_B(t)$ respectively.
The  total Hamiltonian for this system is (setting $\hbar=1$)

\begin{equation}
H_{\rm tot}=\frac{\omega}{2}(\sigma^A_z+\sigma^B_z)+f_A(t)\sigma^A_x+f_B(t)\sigma^B_x.
\end{equation}
This Hamiltonian is familiar in condensed matter theory as an extension of the spin-boson model in the semiclassical regime \cite{SpinBoson}. The classical stochastic
 fields are assumed to obey the following correlation relations in 
the Markov approximation:
\begin{eqnarray}
 &&M[f_A(t)]=M[f_B(t)]=0, \\
 &&M[f_A(t)f_A(s)]=\gamma_A\delta(t-s), \\
 &&M[f_B(t)f_B(s)]=\gamma_B\delta(t-s),\\
 &&M[f_A(t)f_B(s)]=\Gamma\delta(t-s),
\end{eqnarray}
where $M[\cdot]$ denotes the ensemble average over the classical stochastic fields and  for simplicity we 
assume $\gamma_A=\gamma_B=\gamma$.  Obviously,  $\Gamma$ determines 
the respective correlation properties. The effects of the correlation 
between the two fields $f_A(t)$ and $f_B(t)$,  $M[f_A(t)f_B(s)]$, will be revealed in the study of entanglement between the two qubits. 

There has been a lot of work dedicated to  quantum entanglement dynamics in the presence of an environmental noise, 
either in classical or quantum regimes (For some 
recent progress, e.g., see \cite{new-3,new-2,new-20,new-20a,new-1,new0,new1,new2,new3}). Clearly,  having two separated qubits individually coupled to respective external fields would 
reveal the same Lindblad dynamics for each qubit. That is, individual couplings always cause the irreversible decay of entanglement.
 However, when there is a correlation between the two classical stochastic fields, we expect to find new entanglement effects between 
 the two qubits.  It can be shown that the master equation governing the dynamics of two qubits
under the influence of two correlated noises can be derived from the corresponding stochastic Schr\"odinger equation \cite{Yu-etal04},
\begin{eqnarray}
\label{master_eq}
\dot{\rho}&=&-i[H_S, \rho] \\
&&-2\gamma_A(\rho-\sigma^A_x\rho\sigma^A_x)-2\gamma_B(\rho-\sigma^B_x\rho\sigma^B_x) \nonumber \\
 &&-4\Gamma(\sigma^A_x\sigma^B_x\rho+\rho \sigma^A_x\sigma^B_x-\sigma^A_x\rho\sigma^B_x-\sigma^B_x\rho\sigma^A_x). \nonumber
\end{eqnarray}
where $H_S=\frac{1}{2}\omega(\sigma^A_z+\sigma^B_z)$ is the Hamiltonian  of the two qubits.   Upon expanding $\sigma^A_x$ and $\sigma^B_x$ in terms of $\sigma^{A,B}_{\pm}$, we may obtain a more 
solvable form to the master equation:
\begin{eqnarray}
\label{master2}
\dot{\rho}=&-i[H_S, \rho]&-\gamma_A(\sigma^A_+\sigma^A_-\rho-\sigma^A_-\rho\sigma^A_+ - \sigma^A_+\rho\sigma^A_- +\rho\sigma^A_-\sigma^A_+) \\
&& -\gamma_B(\sigma^B_+\sigma^B_-\rho-\sigma^B_-\rho\sigma^B_+ - \sigma^B_+\rho\sigma^B_- +\rho\sigma^B_-\sigma^B_+) \nonumber \\
& & -\Gamma(\sigma^A_+\sigma^B_-\rho-\sigma^B_-\rho\sigma^A_+ - \sigma^A_+\rho\sigma^B_- +\rho\sigma^B_-\sigma^A_+)\nonumber \\
&& -\Gamma(\sigma^B_+\sigma^A_-\rho-\sigma^A_-\rho\sigma^B_+ - \sigma^B_+\rho\sigma^A_- +\rho\sigma^A_-\sigma^B_+) \nonumber +H.C. 
\end{eqnarray} 

At a first glance, Eq.~(\ref{master_eq})  or Eq.~(\ref{master2}) immediately 
displays the dynamics of qubit  A in correlation with field $f_A(t)$, dynamics of qubit B while 
interacting with field $f_B(t)$ and the cross terms due to the correlation between the two fields. Without the involvement of $\Gamma$, the master 
equation would be the sum of two familiar Lindblad master equations for two qubits,
respectively.   In that case, the two qubits would evolve separately throughout time and entanglement between the two qubits will deteriorate with time. 
The correlation between two classical stochastic fields characterized by $\Gamma$ is 
expected to affect entanglement evolution in two different ways dependent on the initial states.  We will show for some initial 
states that entanglement can be significantly enhanced by adjusting the cross correlation between the two fields.  However, for
some other initial states, the cross-correlation works like a catalyst, which may accelerate the decay of entanglement.

\section{Modulated Entanglement Evolution}

Modulated entanglement  evolution will of course depend on the state the system was originally in. If the main goal is to 
try to maintain or improve entanglement,  a good initial state would be that which has maximum entanglement, 
the Bell State, as is presented in cases \ref{bell1}  and \ref{bell2}. It is also interesting to see how a general X type matrix behaves under 
these circumstances, as in case \ref{xform}. 
 It is important to note that $\gamma$ and $\Gamma$ cannot be arbitrary 
and must be chosen in a way that preserves positivity of  the density matrix ($\Gamma \leq \gamma$).

\subsection{Bell State $(|++\rangle,|--\rangle)$}
\label{bell1}
One of the Bell States is described by the pure state vector
\begin{equation}
{|\Psi_0\rangle}=\frac{1}{\sqrt{2}}\{{|++\rangle} + {|--\rangle}\},
\end{equation}
which has the following density operator representation:
\beq
\label{bellstate1}
 \rho=  \frac{1}{\sqrt{2}} \left( \begin{array}{c} 1 \\ 0 \\ 0 \\ 1 \end{array} \right) \frac{1}{\sqrt{2}} \left( \begin{array}{cccc} 1&0&0&1 \end{array} \right) =  \frac{1}{2} \left( \begin{array}{cccc}   1&0&0&1\\  0&0&0&0\\  0&0&0&0\\ 1&0&0&1 \end{array} \right). 
 \eeq

Inserting $\rho$ as the initial reduced density matrix, $\rho(0)$, fuels the following master equation solutions:
\begin{eqnarray*}
\rho_{11}(t) &= & \frac{1}{8\kappa}e^{-(6\gamma+\kappa)t}[2\gamma(1-e^{2\kappa t}) + \kappa(2e^{(6\gamma+\kappa)t}+e^{2\kappa t}+1)] ,\\
\rho_{22}(t) &= & \T \frac{1}{8\kappa}e^{-(6\gamma+\kappa)t}[-2\gamma(1-e^{2\kappa t}) + \kappa(2e^{(6\gamma+\kappa)t}-e^{2\kappa t}-1)], \\
\rho_{23}(t) &= & \T \frac{\Gamma}{\kappa}e^{-(6\gamma+\kappa)t}(e^{2\kappa t}-1), \\
\rho_{14}(t) &= & \T \frac{1}{2}e^{-4\gamma t} ,
\end{eqnarray*} 
where we have defined $\kappa\equiv\sqrt{4\gamma^2+32\Gamma^2}$.  It should be noted that this special initial condition leads to the simple relations:
\[ \begin{array}{ccccccc}
\rho_{33}(t) = \rho_{22}(t) & & \rho_{44}(t)= \rho_{11}(t) & &  \rho_{32}(t)= \rho_{23}(t) & & \rho_{41}(t) =  \rho_{14}(t),\\
\end{array}\]
\[ \begin{array}{cc}
\rho_{12}(t) = & \rho_{13}(t) = \rho_{24}(t) = \rho_{34}(t) = 0, \\
\rho_{21}(t) = & \rho_{31}(t) = \rho_{42}(t) = \rho_{43}(t) = 0.\\
\end{array} \]

Because the off-diagnal elements initially at zero will remain at zero for all time, the time-dependent reduced density operator will be of the X form with only 4 independent, real terms:
\beq
\rho(t)= \left( \begin{array}{cccc}
\rho_{11}(t) & 0 & 0 & \rho_{14}(t)\\
0 & \rho_{22}(t) & \rho_{23}(t) & 0 \\
0 & \rho_{23}(t) & \rho_{22}(t) & 0  \\
\rho_{14}(t) & 0 & 0 & \rho_{11}(t) 
\end{array} \right) .
\eeq

With a solution to the master equation in hand, it is easy to calculate the Concurrence \cite{Wootters}, a measurement of 
entanglement ranging between 0 and 1 evaluated as 
\beq
C(\rho)=2\max\{0, \sqrt{\lambda_1} - \sqrt{\lambda_2} - \sqrt{\lambda_3} - \sqrt{\lambda_4}\},
\eeq
where $\lambda_i$ represent the eigenvalues of the matrix  
\[\varrho=\rho(\sigma^A_y\otimes\sigma^B_y)\rho^*(\sigma^A_y\otimes\sigma^B_y),\]
 in descending order. For this $\rho(t)$, the concurrence throughout time will be 
 \begin{equation}
 C(\rho(t))=2 \max\{0, |\rho_{14}(t)|-\sqrt{\rho_{22}(t)\rho_{33}(t)}\},
 \end{equation} 
 and is plotted  against $\gamma t$ in Fig. 2 for various values of $\Gamma$.
 \begin{figure}[t!]
\begin{center}
\includegraphics[height= 5cm,width=6cm]{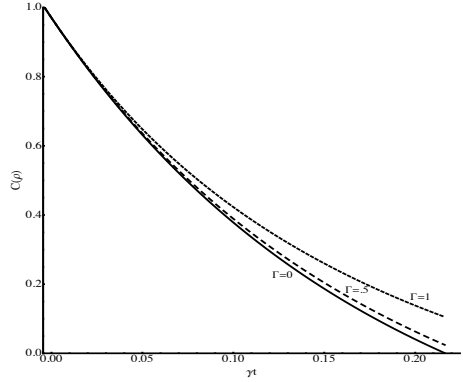}
\caption{Concurrence calculated for various values of $\Gamma$. At times
 later in the decay of the concurrence,  entanglement can be built up by increasing the correlation between the fields.} 
\end{center}
\end{figure}

For all cases of $\gamma$, when $\Gamma=0$, the system exhibits the sudden death of entanglement well-known to the 
local-dephasing channel \cite{Yu-Eberly02}. However, by turning on the correlation $\Gamma$ between the stochastic fields, 
the concurrence curve moves vertically upward, 
denoting an increase in the measurement of 
entanglement. Clearly, we displayed controlled entanglement evolution via correlated environmental noises.
We also are faced with the fact that increasing the correlation $\gamma$ of each atom to their respective fields will also induce 
a faster sudden death of entanglement. By turning 
on the correlation between the fields to its maximum value, $\Gamma=\gamma$, entanglement is maximally enhanced,  as shown in Fig. 2. 

\subsection{Bell State $({|+-\rangle},{|-+\rangle})$}
\label{bell2}
Let's now solve the master equation using the other form of the maximally coherent Bell State:
\begin{equation}
|{\Phi_0}\rangle=\frac{1}{\sqrt{2}}\{{|+-\rangle} + {|-+\rangle}\}.
\end{equation}

The initial reduced density matrix for this state vector is:
\beq
\label{bellstate2}
\rho=  \frac{1}{\sqrt{2}} \left( \begin{array}{c} 0 \\ 1 \\ 1 \\ 0 \end{array} \right) \frac{1}{\sqrt{2}} \left( \begin{array}{cccc} 0&1&1&0 \end{array} \right) =  \frac{1}{2} \left( \begin{array}{cccc}   0&0&0&0\\  0&1&1&0\\  0&1&1&0\\ 0&0&0&0 \end{array} \right) ,
\eeq
which allows us to solve for the time dependent density matrix:
\beqa \begin{array}{cc}
\rho_{12}(t) = & \rho_{13}(t) = \rho_{24}(t) = \rho_{34}(t) = \rho_{14}(t)=0, \\
\rho_{21}(t) = & \rho_{31}(t) = \rho_{42}(t) = \rho_{43}(t) = \rho_{41}(t)= 0. \\
\end{array} 
\eeqa

Using the same definitions as before
\begin{eqnarray*}
\rho_{11}(t) &= & \frac{1}{4\kappa}e^{-(6\gamma+\kappa)t}[-(2\gamma+8\Gamma)(1-e^{2\kappa t}) - \kappa(1+e^{2\kappa t}-2e^{(6\gamma+\kappa)t})] ,\\
\rho_{22}(t) &= & \T \frac{1}{4\kappa}e^{-(6\gamma+\kappa)t}[(2\gamma+8\Gamma)(1-e^{2\kappa t}) +  \kappa(1+e^{2\kappa t}+2e^{(6\gamma+\kappa)t})], \\
\rho_{23}(t) &= & \T  \frac{1}{4\kappa}e^{-(6\gamma+\kappa)t}[-(4\gamma-8\Gamma)(1-e^{2\kappa t}) + 2\kappa(1+e^{2\kappa t})] ,
\end{eqnarray*} 
\[\begin{array}{ccccc}
\rho_{33}(t) =  \rho_{22}(t) & & \rho_{44}(t) = \rho_{11}(t) & & \rho_{32}(t) =  \rho_{23}(t).
\end{array}\]

Making the density operator of a particular X form with 3 independent, real variables throughout all time $t$:
\beq
\label{matrix2}
\rho(t)= \left( \begin{array}{cccc}
\rho_{11}(t) & 0 & 0 & 0 \\
0 & \rho_{22}(t) & \rho_{23}(t) & 0  \\
0 & \rho_{23}(t) & \rho_{22}(t) & 0  \\
0 & 0 & 0 & \rho_{11}(t). 
\end{array} \right) 
\eeq

The concurrence of this density matrix (\ref{matrix2}) is given by:
 \begin{equation}
 C(\rho(t))=2 \max\{0, |\rho_{23}(t)|-\sqrt{\rho_{11}(t)\rho_{44}}\},
 \end{equation}
as plotted in Fig. 3.  
\begin{figure}[t!]
\begin{center}
\includegraphics[width=6.7cm]{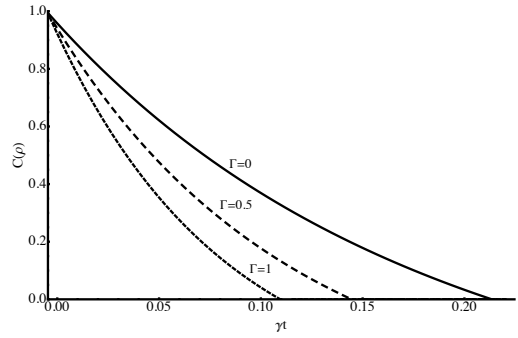}
\caption{Concurrence calculated for various values of $\Gamma$ with $\gamma=1$.  For this initial state, 
 entanglement can only be reduced by turning on the correlation between the fields.} 
\end{center}
\end{figure}

This figure displays quite the opposite of that in the previous section.  Introducing a correlation between the noise fields, $\Gamma$, actually lowers the concurrence curve and 
causes a faster decay of entanglement. This is true for all values of $\gamma$, displaying the impossibility for entanglement enhancement for this initial state evolution. This brings forth 
a very interesting characteristic of entanglement, that even in the Markovian regime where there is no memory of previous times in the system, it is still very sensitive to which initial state 
is being used. Even though all Bell States produce maximum entanglement, one has the prospect for the rebirth of entanglement and the other does not. 

\subsection{X state}
\label{xform}

We can now use a more general approach by utilizing an initial state in the X form which includes the Bell states and Werner states
as special cases \cite{Yu-EberlyQIC07}:
\beq
\rho_x=  \left( \begin{array}{cccc}   a &0&0&w^*\\  0&b&z^*&0\\  0&z&c&0\\ w&0&0&d \end{array} \right) ,
\eeq
 where $a+b+c+d=1$. In this case, the concurrence is 
 \beq
 C(\rho_{\alpha})=2\max\{0, |z|-\sqrt{ad}, |w|-\sqrt{bc}\},
 \eeq
 which allows us to account for a large range of initial entanglement conditions.
 
The solution to the master equation is: 
\begin{eqnarray*}
\rho_{11}(t)&=&\frac{2(a-d)e^{-4\gamma t}+1}{4} + \frac{e^{-6\gamma t}}{4\kappa}[(a-b-c+d)\kappa\cosh{\kappa t}+(16\Gamma z-2\gamma(a-b-c+d))\sinh{\kappa t}],\\
\rho_{22}(t)&=&\frac{2(b-c)e^{-4\gamma t}+1}{4} - \frac{e^{-6\gamma t}}{4\kappa}[(a-b-c+d)\kappa\cosh{\kappa t}+(16\Gamma z-2\gamma(a-b-c+d))\sinh{\kappa t}],\\
\rho_{33}(t)&=&\frac{2(c-b)e^{-4\gamma t}+1}{4} - \frac{e^{-6\gamma t}}{4\kappa}[(a-b-c+d)\kappa\cosh{\kappa t}+(16\Gamma z-2\gamma(a-b-c+d))\sinh{\kappa t}],\\
\rho_{44}(t)&=& \frac{2(d-a)e^{-4\gamma t}+1}{4} + \frac{e^{-6\gamma t}}{4\kappa}[(a-b-c+d)\kappa\cosh{\kappa t}+(16\Gamma z-2\gamma(a-b-c+d))\sinh{\kappa t}],\\
\rho_{23}(t)&=&e^{-6\gamma t}[z\cosh{\kappa t}+\frac{2}{\kappa}((a-b-c+d)\Gamma+\gamma z)\sinh{\kappa t}]=\rho_{32}(t),\\
\rho_{14}(t)&=&we^{-4\gamma t}=\rho_{41}(t).
\end{eqnarray*}

The density operator describing the mixed states of the system will now remain in the X form throughout time with 6 independent, real terms
\beq
\rho(t)= \left( \begin{array}{cccc}
\rho_{11}(t) & 0 & 0 & \rho_{14}(t) \\
0 & \rho_{22}(t) & \rho_{23}(t) & 0  \\
0 & \rho_{23}(t) & \rho_{33}(t) & 0  \\
\rho_{14} & 0 & 0 & \rho_{44}(t) 
\end{array} \right) ,
\eeq
such that the concurrence is calculated as
 \begin{equation}
 C(\rho(t))=2\max\{0, |\rho_{23}(t)|-\sqrt{\rho_{11}(t)\rho_{44}(t)}, |\rho_{14}(t)|-\sqrt{\rho_{22}(t)\rho_{33}(t)} \}.
 \end{equation}
  
The X form initial density matrix displays a more general range of control over the modulation of entanglement. 
 As shown below, the correlation may cause entanglement to decay faster initially, but it can compensate the loss of 
 entanglement later. To see this, we choose specific values of the initial 
X form density matrix
 \beq
\rho(0)= \frac{1}{3}\left( \begin{array}{cccc}
\frac{1}{2} & 0 & 0 & \frac{3}{2} \\
0 & 1 & 1 & 0  \\
0 & 1 & 1 & 0  \\
\frac{3}{2} & 0 & 0 & \frac{1}{2} 
\end{array} \right) 
\eeq
and plot the concurrence through time in Fig. 4.
\begin{figure}[t!]
\label{fig05}
\begin{center}
\includegraphics[width=6cm]{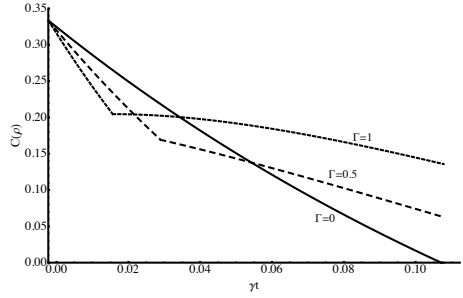}
\caption{Concurrence plotted over $\gamma t$ for various values of $\Gamma$. At first entanglement is reduced due to correlated noise but later begins to enhance it.} 
\end{center}
\end{figure}
 
The results demonstrate that the entanglement can be significantly enhanced by the cross-correlation at late times even though the correlation was detrimental to entanglement 
at early times. More explicitly,  it can be shown that at early time $|\rho_{23}(t)|-\sqrt{\rho_{11}(t)\rho_{44}(t)}$ is the dominant term in the calculation of concurrence, resulting in the 
degradation of  entanglement due to the cross-correlation. However, at a certain point $|\rho_{14}(t)|-\sqrt{\rho_{22}(t)\rho_{33}(t)}$ begins to dominate, causing the correlation to 
enhance the entanglement. This demonstrates our ability to improve entanglement for a wide range of initial states and time scales.

\section{Conclusion}
In summary, we studied a system of two separated qubits each coupled to a stochastic field which can then be correlated, opening the option for the
 enhancement or reduction of entanglement between the qubits. By solving the master equation for the qubit dynamics under various initial states and viewing 
 the concurrence as a function of time, the effects of the correlation between the fields, $\Gamma$, became imminent. Of our three cases for initial 
 states, the Bell State, Eq. (\ref{bellstate1}), showed a potential for the regeneration of entanglement.  The amount of 
 entanglement this particular case can achieve is dependent on the degree of correlation between the two noises. 
In the second case we considered the Bell State of Eq. (\ref{bellstate2}) for which correlation $\Gamma$ did 
not cause the entanglement between the atoms to be regenerated. In fact, it instead reduced the entanglement and caused it to decay at an even faster rate. This 
illuminates a selectivity that entanglement has toward the initial state of the system, even under the memoryless Markov approximation. It is important then to look
 at the general X form density matrix  of Section \ref{xform} allowing us to account for various initial entanglement conditions of the system. This case presented the ability to
 modulate entanglement in either way. The range of maximum enhancement was variable depending on the parameters of the initial matrix, giving us a wide range 
 of control over the improvement of entanglement.

In general, many characteristic properties of entanglement are still unknown. In this model system, we have created a bridge 
from one qubit to the other through the correlation of the stochastic fields, allowing for entanglement to be modulated.  This correlated noise approach can easily
 be applied to multipartite systems, which in turn might provide an even higher enhancement of entanglement \cite{high-1,high0,high1,high2}. 
 Finally, it may be worth noting that the relationship between the classical noise model presented here and the fully quantized models discussed in \cite{FicTan06,ASH}
 is an interesting problem that will be addressed in future publications.

\section{Acknowledgements}
We acknowledge partial  financial support from the following agencies: DARPA HR0011-09-1-0008 and NSF PHY-0925174.
B.C. also acknowledges Stanley Fellowship support from Stevens Institute of Technology. 



\end{document}